\documentclass[conference,emptycopyrightspace]{IEEEtran}
\IEEEoverridecommandlockouts
\usepackage{cite}
\usepackage{amsmath,amssymb,amsfonts}
\usepackage{algorithmic}
\usepackage{graphicx}
\usepackage{textcomp}
\usepackage{xcolor}
\usepackage{comment}
\usepackage{subfig}
\usepackage{bm} 

\begin{document}

\title{Secure Multi-hop Telemetry Broadcasts \\ for UAV Swarm Communication
\thanks{This research has been funded partly by the Federal Ministry of Education and Research of Germany under grant numbers 16ES1131 and 16ES1128K. The authors are responsible for the content of this publication.}
}

\author{
\IEEEauthorblockN{Randolf Rotta$^1$, Pavlo Mykytyn$^1$$^,$$^2$}
\IEEEauthorblockA{$^1$Brandenburg University of Technology, Cottbus, Germany\\
$^2$IHP - Leibniz Institute for High-Performance Microelectronics, Frankfurt (Oder), Germany}
}

\maketitle

\begin{abstract}
Unmanned Aerial Vehicles (UAVs) are evolving as adaptable platforms for a wide range of applications such as precise inspections, emergency response, and remote sensing. Autonomous UAV swarms require efficient and stable communication during deployment for a successful mission execution.
For instance, the periodic exchange of telemetry data between all swarm members provides the foundation for formation flight and collision avoidance.
However, due to the mobility of the vehicles and instability of wireless transmissions, maintaining a secure and reliable all-to-all communication remains challenging. This paper investigates encrypted and authenticated multi-hop broadcast communication based on the transmission of custom IEEE 802.11 Wi-Fi data frames.

\end{abstract}


\section{Introduction}

To share data with a ground control station (GCS), other UAVs in a swarm, or other centralized infrastructure, UAVs rely on wireless communication. UAVs communicate with the base station to receive commands and transmit sensor data. Such point-to-point communication between a single vehicle and a GCS is straightforward. 

However, when considering a swarm with multiple vehicles and ground stations, enabling efficient and secure communication becomes increasingly challenging. The communication security is critical because the swarm relies on it for cooperation, formation forming, or collision avoidance. However, wireless communications is particularly vulnerable to intentional and unintentional interference, jamming, interception, eavesdropping, and enables cyber-attacks targeting data privacy and integrity \cite{003}. Furthermore, the dynamic nature of UAV networks with nodes continuously moving and re-establishing the connection between one another makes dependable communication links much more difficult to maintain. The dynamic and mobile nature of UAVs, combined with the limited range of wireless communication, necessitates the use of multi-hop communication techniques.

In this paper we focus on security and reliability aspects of all-to-all multi-hop broadcast communication between UAV swarm members and the GCS.
Low-cost microprocessors with a few megabytes of RAM and integrated Wi-Fi radio open up the opportunity for new experiments with mesh protocols. One example is the Espressif ESP32 series, which allows broadcasting and receiving custom Wi-Fi frames without the need to associate to any Access Point. Securing these transmissions is left to be the protocol implementation.

\section{Related work}

Mesh communication protocols such as B.A.T.M.A.N and Babel are often considered as a basis for UAV swarms~\cite{app11104363}. Their design assumes that security mechanisms are handled on higher layers. 
Because of this assumption, mesh protocol layers are open to a variety of cyber-attacks. The authors in \cite{001} mention that in multi-hop UAV swarm communication, active RF jamming and eavesdropping are among the most common cyber-attacks. Additionally, UAV swarms using the Robot Operating System (ROS 1) are vulnerable to a variety of cyber-attacks. ROS 2, on the other hand, has eliminated some of the issues mentioned above by introducing authentication and encryption based on a public key infrastructure.

Babel~\cite{rfc8966} proposes two optional mechanisms based on shared keys or Datagram Transport Layer Security (DTLS). SecBATMAN~\cite{racz2013security} proposes a security extension but lacks a dynamic key exchange. Studies like~\cite{6952779} argue, that dynamic key management schemes are necessary in order to secure these mesh protocols.
Beyond mesh protocols that try to optimize forwarding routes through the network, Synchronous Flooding (SF) provides a much simpler alternative, c.f.~\cite{10.1145/3410159}. These protocols synchronize the forwarding of broadcasts such that a $n$-hop broadcast needs just $n$ consecutive time slots independent of the actual number of participating nodes.
Unfortunately, existing implementations focus on Bluetooth Low Energy and IEEE 802.15.4 radios, which limits the available throughput. To our best knowledge no implementation based on IEEE 802.11 exists.

\section{Multi-hop Telemetry Broadcasts}

Point-to-point communication refers to a direct communication link established between two devices, such as an individual UAV and the GCS. Multi-point communication in UAV swarms connects a single GCS to multiple UAVs and multi-point meshes add multi-hop routing between GCS and UAVs.
While this link to the ground provides essential connectivity, it is insufficient to facilitate seamless data exchange within the swarm.
Mesh protocols like B.A.T.M.A.N and Babel focus on providing multi-hop point-to-point communication between arbitrary network nodes. This would be perfect, for example, for two vehicles that cooperate on a task. 
However, cooperation between multiple vehicles requires to broadcast at least their position telemetry frequently enough to all the other UAVs.
Flooding the mesh network with each vehicle's telemetry would work, but is inefficient because it re-broadcasts more often than necessary. This reduces the available throughput and increases the risk of colliding transmissions.
Therefore, efficient multi-hop broadcast mechanisms are needed. 
We propose to revisit the flooding of route discovery messages in proactive mesh protocols like B.A.T.M.A.N and Babel. They use flooding to learn the best path to each possible destination node. Only the best next hop towards each destination node is stored, which is repeated at the next node until the destination node is reached. This approach provides an advantage of adjusting the path, while messages are already traveling. The next hops essentially form a collection tree toward each destination node. Our approach aims at inverting these trees into broadcast trees originating from that node. Thus, together with each outgoing message the node broadcasts its next hop table. The receiver of that message can then figure out its position in each broadcast tree and avoid unnecessary transmissions. This information allows each node to selectively forward pending messages from the queue based on the neighbor's needs. 
Typically, telemetry messages in the MAVLink protocol are much smaller than 256 bytes. Thus, during forwarding, multiple messages from different sources can be aggregated into a single IEEE 802.11 frame.
A similar pattern can be achieved with ROS2-based communication. The underlying data distribution service allows to configure forwarding of published messages, for example, to multicast IP addresses. It can also be configured to receive such multicast messages. However, the Real-time Publish-Subscribe Protocol (RTPS) that is used between the ROS2 nodes is much more complex than MAVLink.
Instead of maintaining a separate broadcast tree rooted at each node, a single spanning tree could be sufficient~\cite{juttner2005tree}. A broadcast message is re-broadcasted only in case if it was received via a neighbor in the spanning tree, it is not a leaf node, and the message was not re-broadcasted before. 
This approach combined with the loop avoidance techniques of the Babel protocol~\cite{rfc8966} should generate even more effective results.

\section{Security in UAV communication}

UAV swarm communication is susceptible to RF jamming, Man-in-the-Middle (MITM), Eavesdropping, Traffic Analysis (TA), and Replay attacks. Eavesdropping and TA are passive cyber-attacks and require additional hardware in order to detect them. However, by integrating data encryption mechanisms, the effects of these cyber-attacks can be mitigated. The effects of a MITM attack, can be mitigated by integrating data authentication mechanisms. The effects of a Replay attack can be mitigated by integrating a timestamp to deem old and repeated messages invalid.
Our approach is based on providing a secure and authenticated communication for all of the UAV swarm members. As a key exchange protocol we plan to use the Elliptic-curve Diffie–Hellman (ECDH) adapted to be used for multiple parties. Once all of the parties have generated their private and public keys and calculated a common shared secret (session key), we will authenticate the message using Hash-based Message Authentication Code (HMAC) based on Secure Hashing Algorithm (SHA-2) by generating a hash of the message together with the session key and appending the first 16 bytes of the hash to the end of the message, thus providing authentication of the contents of the broadcasted message. To encrypt the broadcasted message we will utilize the Advanced Encryption Standard (AES) with the 128-bit key. Each sent message will also include a timestamp to protect against Replay attacks and deem old messages invalid.
By using this approach, the UAV swarm members can establish a common shared secret through ECDH key exchange, enabling secure and authenticated communication within the UAV swarm.

\section{Conclusions}


This paper has presented an idea for multi-hop telemetry broadcasts communication within a UAV swarm, specifically designed to enable fast, efficient, and secure mesh communication for mission execution and collision avoidance purposes. The proposed approach incorporates all-to-all broadcasts using flooding and message relaying. 
By leveraging the ECDH group key exchange protocol, drones establish a shared secret, ensuring secure communication channels within the mesh network. The use of AES-128 encryption guarantees the confidentiality of telemetry broadcasts, protecting sensitive information from unauthorized access.
To ensure message integrity and authenticity, each broadcast message includes a HMAC-256 signature with a timestamp. This signature provides a reliable means to verify the origin and integrity of the message, preventing tampering or spoofing attempts.

\bibliographystyle{IEEEtran}
\bibliography{references}

\end{document}